\documentclass{nature}

\usepackage{graphicx}
\usepackage{latexsym}
\usepackage{amssymb}
\usepackage{amsmath}
\usepackage{amsfonts}
\usepackage{gensymb}
\usepackage{bm}
\usepackage{braket}
\usepackage{physics}
\usepackage{bbold}
\usepackage{cases}
\usepackage{tabularx}
\usepackage{multirow}
\usepackage{braket}
\usepackage{epstopdf}
\usepackage[urlcolor=blue]{hyperref}

\newcommand{\be}{\begin{equation}}
\newcommand{\ee}{\end{equation}}

\newcommand{\bea}{\begin{eqnarray}}
\newcommand{\eea}{\end{eqnarray}}

\usepackage{color}
\usepackage{dcolumn}
\usepackage{epsfig}
\usepackage{bm}
\usepackage[urlcolor=blue]{hyperref}
\hypersetup{backref, colorlinks=true, linkcolor=blue, citecolor=blue}

\title{Critical field measure for topological superconductivity}

\author{Noah F. Q. Yuan$^{1,\dagger}$ and Xiao-Jia Chen$^{1,2,\star}$}

\begin{document}

\maketitle

\begin{affiliations}
\item{School of Science, Harbin Institute of Technology, Shenzhen 518055, China}
\item{Center for High Pressure Science and Technology Advanced Research, Shanghai 201203, China}
$^{\dagger}$e-mail: fyuanaa@connect.ust.hk,\quad
$^{\star}$e-mail: {xjchen2@gmail.com}
\end{affiliations}

\begin{abstract}
A promising direction for harnessing the laws of quantum mechanics to perform quantum computation\cite{chuang} is the topological quantum computation\cite{Kitaev,nayak}, for which topological superconductivity is one of the physical platforms. Intensive theoretical studies\cite{ssasa,kitaev,read-green,ivanov,FuKane,xlqi,Alicea,qixl,law,flensberg,beenakker} have been carried out and followed by tremendous experimental efforts\cite{mour,nadj,jpxu,hhsun,agga,hwang,pzhang,dwang,szhu,lim,jack,yuanyh,kezi,nayakak,Stanene} in the realization of topological superconductivity, though debates remain mainly because of the inadequacy of the convincing detection techniques\cite{qlhe,sasa,peng,kayy,jliu,DaSarma}.
Here we report a theoretical finding in a superconductor with surface or edge states,
where the critical field is found to obey the unique power-law temperature dependence $B_{c3}\sim (T_c-T)^{\gamma}$ with $T_c$ being the onset critical temperature of superconductivity and the fractional exponent $\gamma= 2/3$, differing from the conventional values of $\gamma=1/2$ and $1$. 
The topological surface superconductivity is hence expected in the three-dimensional (3D) system and the topological edge superconductivity in the two-dimensional (2D) system. The application of this measure to the intrinsic topological superconductors FeTe$_{1-x}$Se$_{x}$ ($x\sim 0.5$) supports the 
validity of such an accessible, convenient, and reliable transport technique for the identification of topological superconductivity. The new form of theory on topological superconductivity together with the developed identification technique is expected to guide the search of topological superconductors in future. 
\end{abstract}

At the dawn of the 21st century, Alexei Yu Kitaev proposed a one-dimensional (1D) theoretical model of unconventional $p$-wave superconductor\cite{kitaev}, where exotic topological excitations known as Majorana zero modes could arise at its two ends. During the same period of time, N. Read and Dmitry Green found the 2D counterpart model of $p$-wave superconductor\cite{read-green}, which could host Majorana zero modes at the vortex cores. Majorana zero modes in such topological superconductors are robust against disorders and obey the non-Abelian statistics, thus can be used in topological quantum computations\cite{ivanov,Alicea}. Both the Kitaev chain model and the Read-Green model require $p$-wave superconductivity, which is rare in nature. For a long time, Sr$_2$RuO$_4$ had been considered as the only candidate of $p$-wave superconductor\cite{rice,mackenzie}. However, recent data of nuclear magnetic resonance has ruled out the possibility of $p$-wave pairing in Sr$_2$RuO$_4$\cite{pustogow}.

Inspired by the Kitaev chain model and the Read-Green model, more theoretical models of topological superconductivity are developed in conventional $s$-wave superconducting systems\cite{FuKane,qixl,xlqi,Alicea}. In 3D, the heterostructure formed by an $s$-wave superconductor (SC) and a topological insulator (TI) was proposed by Liang Fu and Charles L. Kane\cite{FuKane}, where
topological superconductivity is expected if superconductivity and topological surface states coexist in the SC-TI interface\cite{jpxu,hhsun}. 
In 2D, the heterostructure formed by an $s$-wave SC and a quantum anomalous Hall (QAH) system was proposed by X.-L. Qi, Taylor L. Hughes and Shou-Cheng Zhang\cite{qixl,xlqi}, where topological superconductivity is expected if superconductivity and topological surface states coexist in the SC-QAH interface. Thus, in order to engineer topological superconductivity in a heterostructure\cite{kezi,jack}, one first needs to ensure the coexistence of superconductivity and topological states in its interface.
On the other hand, it was recently reported that in some 3D\cite{agga,hwang,pzhang,dwang,szhu,lim,yuanyh} and 2D\cite{nayakak,Stanene} materials, superconductivity and topological states may coexist, where topological superconductivity could take place. In these intrinsic candidates of topological superconductors, one also needs to confirm that the superconductivity and topological states do coexist on the open surface (3D) or along the open edge (2D). As such, the identification of superconductivity coexisting with topological states on the boundaries (interface, common edge, open surface and open edge) has been becoming a critical issue in the exploration of topological superconductivity. In the following, we demonstrate that the nucleation of superconductivity\cite{vortex,deGennes,Abrikosov,tinkham,saint,saint-james,Khly,Matsumoto} could provide an answer to this question.

Nucleation is the first step of a thermodynamic phase transition prior to the bulk transition when there are preferential sites. In our case of topological superconductors, boundaries are preferential for the Cooper pairs to accumulate. Deep in the superconducting phase, contributions from the boundaries are negligible due to the lower dimensionality compared to the bulk. However, near the bulk phase transition, the bulk coherence length becomes so divergent that boundaries can also affect the bulk significantly, resulting in a nucleation prior to the bulk phase transition. 

To explore the nucleation of topological superconductivity on surfaces, similar to the previous theoretical models, we construct a theoretical model in Fig. 1\textbf{a}, where the 3D superconductor is in the semi-infinite region $z\leq 0$, with topological surface states near the open surface $z=0$, and topological surface superconductivity is realized.

We consider the Ginzburg-Landau (GL) regime, which is near the superconducting phase transitions and at distances greater than the zero-temperature bulk coherence length $\xi_0$. 
Since the topological surface states usually extend to the bulk by a few atomic layers\cite{TI}, on the scale of $\xi_0$ we can assume that the topological surface states only live on the 2D open surface $z=0$. After washing out the short-range details, the GL free energy is composed of the bulk and surface terms
\be\label{eq_F}
    F=\int_{z\leq 0} d^3\bm r\left\{\frac{1}{2m}\left|\bm D\psi\right|^2+\mathcal{A}|\psi|^2\right\}+\sigma\int_{z=0} d^{2}\bm r |\psi|^2,
\ee
where $\psi(\bm r)$ is the superconducting order parameter 
and $\bm r=(x,y,z)$ is the spatial coordinate.

The bulk Cooper pairs are described by the kinetic energy $|\bm D\psi|^2/2m$ and the potential energy $\mathcal{A}|\psi|^2$. Here $m$ is the electron mass, $\bm D=-i\nabla-2\pi\bm A/\Phi_0$ is the momentum operator with vector potential $\bm A$ and flux quantum $\Phi_0=h/(2e)$, and $\mathcal{A}=\alpha(T-T_{c0})$ depends on temperature $T$ with $T_{c0}$ being the bulk critical temperature. The surface Cooper pairs are described by the surface energy $\sigma$, which can be negative, zero and positive.
In our model, without topological surface states, the surface energy is zero; while with topological surface states, the surface energy is negative.
Microscopic derivations of such conclusion and Eq. (\ref{eq_F}) can be found in Methods.
We would like to point out that $\psi$ describes the pairing correlation, which is related to but not directly proportional to the energy gap of Bogouliubov quasiparticles.

In the maintext, we deal with the topological case of negative surface energy ($\sigma<0$). 
Other cases of $\sigma\geq 0$ can be found in Methods. 

First we consider the nucleation at zero field. The free energy Eq. (\ref{eq_F}) has a negative surface energy, suggesting the preferential accumulation of the Cooper pairs on the open surface. As a result, the superconducting phase transition at zero field consists of two steps as the temperature goes down, whose order parameters are plotted in Fig. 1\textbf{b}. As shown in Fig. 1\textbf{b}, the first step (blue line) is analogous to the heterogeneous nucleation in the vapor to liquid condensation, where the electron gas condenses into Cooper pairs on the open surface, while the bulk electrons stay mostly normal. This step is at temperatures above the bulk critical temperature $T>T_{c0}$, whose the order parameter is mainly localized on the surface layer. In the second step (black line) when the temperature is lower than the bulk critical temperature $T<T_{c0}$, besides the surface layer, the bulk is also superconducting. In both cases, the extrapolations of the order parameters (dashed lines) intercept the $z$-axis at the same point, the de Gennes extrapolation length\cite{deGennes}, given by $l_s\equiv(2m|\sigma|)^{-1}$. We have shown that, at zero field the negative surface energy requires the order parameter to increase in the surface layer of thickness $\sim l_s$, so that the nucleated superconductivity can bear stronger thermal fluctuations, and the onset critical temperature $T_c=T_{c0}+2m\sigma^2/\alpha>T_{c0}$ is higher than the bulk one. 

Next we consider the nucleation under a magnetic field $\bm B=\nabla\times\bm A$. In fact, the mechanism at zero field also applies to the nucleation under finite fields, in which the negative surface energy requires order parameters to increase near the surface, so that the nucleated superconductivity can bear stronger magnetic fields. This enhanced critical field is denoted as the nucleation critical field $B_{c3}(T)$. 
In this work we focus on the type-II superconductor, but our theory of nucleation critical field also applies to a type-I superconductor sufficiently close to $T_c$, as long as its Ginzburg parameter is not too small\cite{Khly}.

We start with examining the bulk nucleation of type-II superconductors under finite fields. Near the upper critical field $B_{c2}$, any bulk line away from the open surface could be a nucleus for Cooper pairs to accumulate, as found by Abrikosov in terms of vortices\cite{vortex}. When the bulk nucleus moves to the surface region, the negative surface energy requires an increase in the order parameter near the surface, which stabilizes the surface superconductivity and hence leads to $B_{c3}>B_{c2}$.

Due to the isotropic bulk free energy in Eq. (\ref{eq_F}), the upper critical field $B_{c2}$ of the bulk is isotropic. However, the nucleation critical field $B_{c3}$ is highly anisotropic, due to the negative surface energy of the $z=0$ open surface instead of other surfaces. 

We first consider in-plane fields $\bm B\parallel xy$-plane. 
As shown in Fig. 2\textbf{a}, the bulk nucleus is Gaussian along the $z$-axis under an in-plane field $B\sim B_{c2}$. As $B$ increases the nucleus moves to the surface region, where the order parameter is enhanced compared with the bulk nucleus (dashed line), resulting in the surface superconductivity.

At weak fields and temperatures close to $T_c$, the nucleated surface superconductivity is essentially 2D in the $xy$ layer with an effective thickness $\sim l_s$ along the $z$-axis. The nucleation critical field is thus $B_{c3}={\Phi_0}/{(\pi\xi l_s)}\propto(T_c-T)^{1/2}$ with the coherence length $\xi=(2m\alpha|T_c-T|)^{-1/2}$, and the critical exponent of the in-plane magnetic field is 1/2\cite{deGennes,tinkham,Abrikosov,Matsumoto,Khly}.

As the magnetic field increases and the temperature is slightly away from $T_c$, the nucleation critical field is found to be $B_{c3}\propto |T_c-T|^\gamma$ with the fractional exponent $\gamma=2/3$, based on a heuristic argument as follows. 
In the trivial case, the inhomogeneity of nucleated bulk superconductivity along the $z$-axis is characterized by the coherence length $\xi$, and when the magnetic flux trapped in the area $\xi^2$ is on the order of $\Phi_0$, the nucleated bulk superconductivity is destroyed.
In the topological case, the inhomogeneity of nucleated surface superconductivity along the $z$-axis turns out to be characterized by the length $\zeta=(\xi^2 l_s)^{1/3}$. Similarly, when the magnetic flux trapped in the area $\zeta^2$ is on the order of $\Phi_0$, the nucleated surface superconductivity is destroyed. 
Hence the in-plane nucleation critical field in the intermediate field regime is
\be\label{eq_Bxy}
B_{c3}^{\rm in}=0.525\frac{\Phi_0}{\left(\xi^2 l_s\right)^{\frac{2}{3}}}\propto(T_c-T)^{2/3}.
\ee

The exact order parameter and the exact nucleation critical field in the full GL regime are given in Methods, from which the weak field regime is found as $B<0.01B_s$, and the intermediate field regime is $0.1B_s<B<B_s$ with $B_s\equiv \Phi_0/(\pi l_s^2)$.

The full phase diagram of temperature and in-plane magnetic field in the GL regime is shown in Fig. 2\textbf{c}. 
As field increases, the bulk superconductivity stays uniform until the lower critical field $B_{c1}$ where vortices start to nucleate, and the bulk is in the mixed phase when $B_{c1}<B<B_{c2}$.
In the regime $B_{c2}<B<B_{c3}$, nucleated superconductivity and topological states coexist on the surface, resulting in the \emph{topological surface superconductivity} (TSSC).

Then we consider out-of-plane fields $\bm B\parallel z$-axis. As shown in Fig. 2\textbf{b}, the bulk nucleus is uniform along the $z$-axis under an out-of-plane field $B\sim B_{c2}$. As $B$ further increases, near the open surface $z=0$ superconductivity nucleates due to negative surface energy, resulting in the surface superconductivity. 
The out-of-plane nucleation critical field is similar to the one obtained by Abrikosov\cite{vortex}
\be\label{eq_Bz}
B_{c3}^{\rm out}=\frac{\Phi_0}{2\pi\xi^2}\propto(T_c-T)^{1}.
\ee
The full phase diagram of temperature and out-of-plane magnetic field in the GL regime is shown in Fig. 2\textbf{d}. 
We find the same lower and upper critical fields $B_{c1}$ and $B_{c2}$ as those of the in-plane case, while $B_{c3}$ can be obtained by parallel shifting of $B_{c2}$ along the temperature axis from $T_{c0}$ to $T_c$.
TSSC is also found in the regime $B_{c2}<B<B_{c3}$.

We have thus found two types of nuclei in 3D topological superconductors. At zero field, the only nucleus is the surface, and under finite fields, bulk nucleus also arises due to the orbital effect.  
Our methods can also be applied in 2D topological superconductors where topological surface states are replaced by topological edge states, and the surface nucleus is replaced by the edge nucleus. The model and results are summarized in Fig. 3. 

We construct a 2D theoretical model as shown in Fig. 3\textbf{a}, where the 2D superconductor is confined in the 2D plane of $z=0$, topological edge states exist along the edge $x=0$, and bulk superconductivity in the semi-infinite region $x\leq 0$. Similar to the 3D case, at zero field the superconducting phase transition consists of two steps as the temperature goes down, whose order parameters are plotted in Fig. 3\textbf{b}.

Under in-plane fields, the orbital effect is screened\cite{tinkham} and hence only edge nucleus remains as shown in Fig. 3\textbf{c}. In the resulting phase diagram Fig. 3\textbf{e}, the mixed phase and the lower critical field $B_{c1}$ caused by the orbital effect are absent, and the TSSC is replaced by the \emph{topological edge superconductivity} (TESC), where the nucleated superconductivity and topological states coexist along the edge. Importantly, both the upper critical field $B_{c2}\propto\sqrt{T_{c0}-T}$ and the nucleation critical field $B_{c3}\propto\sqrt{T_c-T}$ follow the square-root behaviour.

Under out-of-plane fields, both bulk and surface nuclei are present as shown in Fig. 3\textbf{d}. The resulting phase diagram Fig. 3\textbf{f} is the same as that of the 3D superconductor under in-plane fields, where the upper critical field $B_{c2}\propto {T_{c0}-T}$ and the nucleation critical field $B_{c3}\propto({T_c-T})^{2/3}$ follow different power laws.
 
In Table 1 we list the critical exponent $\gamma$ of critical fields along the in-plane and out-of-plane directions for both the topological and trivial cases, where $\gamma$ is defined by $B_{c3}\sim (T_c-T)^\gamma$, and the superconductor can be 3D or 2D. 
Detailed derivations of Table 1 can be found in Methods.

Our theory could be applied in intrinsic topological superconductors\cite{agga,hwang,pzhang,dwang,szhu,Stanene,lim,yuanyh,nayakak}, or superconducting heterostructures\cite{jpxu,hhsun,jack,kezi}, where
the topological states can be inherited from the normal phase\cite{agga,hwang,pzhang,dwang,szhu,Stanene,lim,yuanyh,jpxu,hhsun,jack,kezi}, or emerge in the superconducting phase\cite{nayakak}. 

To examine the validity of the proposed critical field measure from experiments, we choose the well defined candidate of intrinsic topological superconductors FeTe$_{1-x}$Se$_{x}$ ($x\sim 0.5$) as examples. The theoretical calculations\cite{sczhang} revealed the topological surface states on the (001) surface in such systems. The substitution composition $x$ around 0.5 was found to be critical for holding both the superconducting and topological states\cite{neutron}. The predicted topological superconductivity was indeed detected for FeTe$_{0.55}$Se$_{0.45}$ from the high-resolution spin-resolved and angle-resolved photoelectron spectroscopy measurements\cite{pzhang}. The experimental evidence for Majorana bound states was further provided on the same superconductor by using scanning tunneling spectroscopy on the superconducting Dirac surface state\cite{dwang,kong}. Thus, FeTe$_{1-x}$Se$_{x}$ ($x\sim 0.5$) system provides an excellent platform for the realization of the Majorana bound states and for the examination of any theories of topological superconductivity. 

Thanks for the high magnetic field measurements which were carried out on FeTe$_{0.52}$Se$_{0.48}$ and FeTe$_{0.5}$Se$_{0.5}$ at Toulouse in 2010\cite{res} and 2014\cite{tdo}, respectively. The upper critical fields, collected from the measurements of the resistance\cite{res} and tunnel diode oscillator frequency\cite{tdo}, can be used to compare with the present theoretical prediction for the magnetic field directions in the $ab$ plane and along the $c$ axis, respectively. For 3D topological superconductors, the critical exponent $\gamma$ should be $2/3$ and $1$ for the applied magnetic field direction in the $ab$ plane and along the $c$ direction, respectively, over the wide regime of intermediate magnetic fields (see Table 1). As shown in Fig. 4, the theory reproduces very well the experimental results for FeTe$_{0.52}$Se$_{0.48}$\cite{res} and FeTe$_{0.5}$Se$_{0.5}$\cite{tdo} in two magnetic field directions. The excellent agreement between the theory and experiments adds solid evidence for the topological superconductivity in the studied FeTe$_{1-x}$Se$_{x}$ from the high magnetic field measurements. Importantly, the comparison demonstrates that the critical field measure is indeed effective and powerful in judging topological superconductivity. 

Besides the critical field measure, the electrical resistivity at zero field also carries up the feature of topological superconductors. 
For 2D topological superconductors, one can measure the local tunneling conductance at zero bias\cite{nayakak}, which should resemble Fig. 1\textbf{b} in the topological case. 
For 3D topological superconductors, one can analyze the evolution of electrical resistivity during the superconducting phase transition.
According to the Mermin-Wagner theorem\cite{Hohenberg,Mermin-Wagner,Coleman}, long-range fluctuations are favored in dimensions equal to or less than 2. 
Hence in the clean limit of the trivial case, the superconducting transition is signaled by a sharp drop of the resistivity at the critical temperature for 3D superconductors, while a slow resistivity transition due to fluctuations is expected for 2D superconductors. In a 3D topological superconductor, the nucleation on the 2D surfaces is the first step prior to the 3D bulk transition, and as a result, a slow resistivity transition from the onset superconducting state (surface superconductivity at $T_c$) to the complete zero-resistivity superconducting state (bulk superconductivity at $T_{c0}$) should be found, similar to the trivial 2D case instead of the trivial 3D case.
The slow resistivity transition with a relatively large width is the additional transport signature for the topological superconductivity in a 3D material even perfectly grown.

\begin{addendum}

\item[Acknowledgements] N.F.Q.Y. acknowledges the National Natural Science Foundation of China (Grant. No. 12174021) for the financial support. X.J.C. thanks the financial support from the Shenzhen Science and Technology Program (Grant No. KQTD20200820113045081), the Basic Research Program of Shenzhen (Grant No. JCYJ20200109112810241), and the National Key R$\&$D Program of China (Grant No. 2018YFA0305900).  

\end{addendum}

\newpage
\begin{center}
\includegraphics[width=\columnwidth]{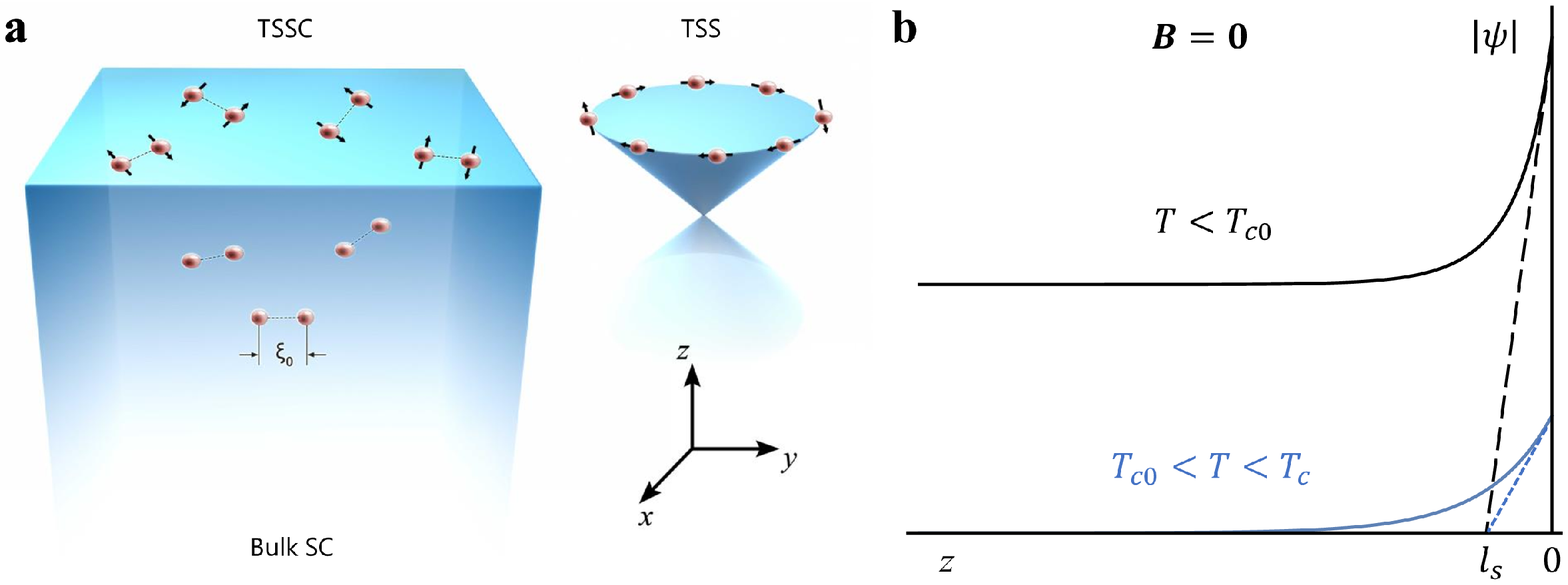}
\end{center}
\vspace{-0.5cm}
\noindent\textbf{Figure 1 $\mid$ Nucleation at zero field in the 3D topological superconductor}. \textbf{a}, The 3D topological superconductor with bulk superconductivity (Bulk SC) in the semi-infinite region $z\leq 0$ and topological surface superconductivity (TSSC) near the open surface $z=0$. TSSC is the result of the interplay between bulk superconductivity and topological surface states (TSS). Here $\xi_0$ is the zero-temperature coherence length of the bulk, and in this work we consider long-range physics at distances greater than $\xi_0$. \textbf{b}, Surface nucleation at zero field. The blue and black lines correspond to the order parameters in the temperature regimes $T_{c0}<T<T_c$ and $T<T_{c0}$, respectively. Dashed lines denote the extrapolations of order parameters, which intercept the $z$-axis at the extrapolation length $l_s$. The nucleus of the order parameter is the surface $z=0$, as depicted in \textbf{a}.

\newpage
\begin{center}
\includegraphics[width=\columnwidth]{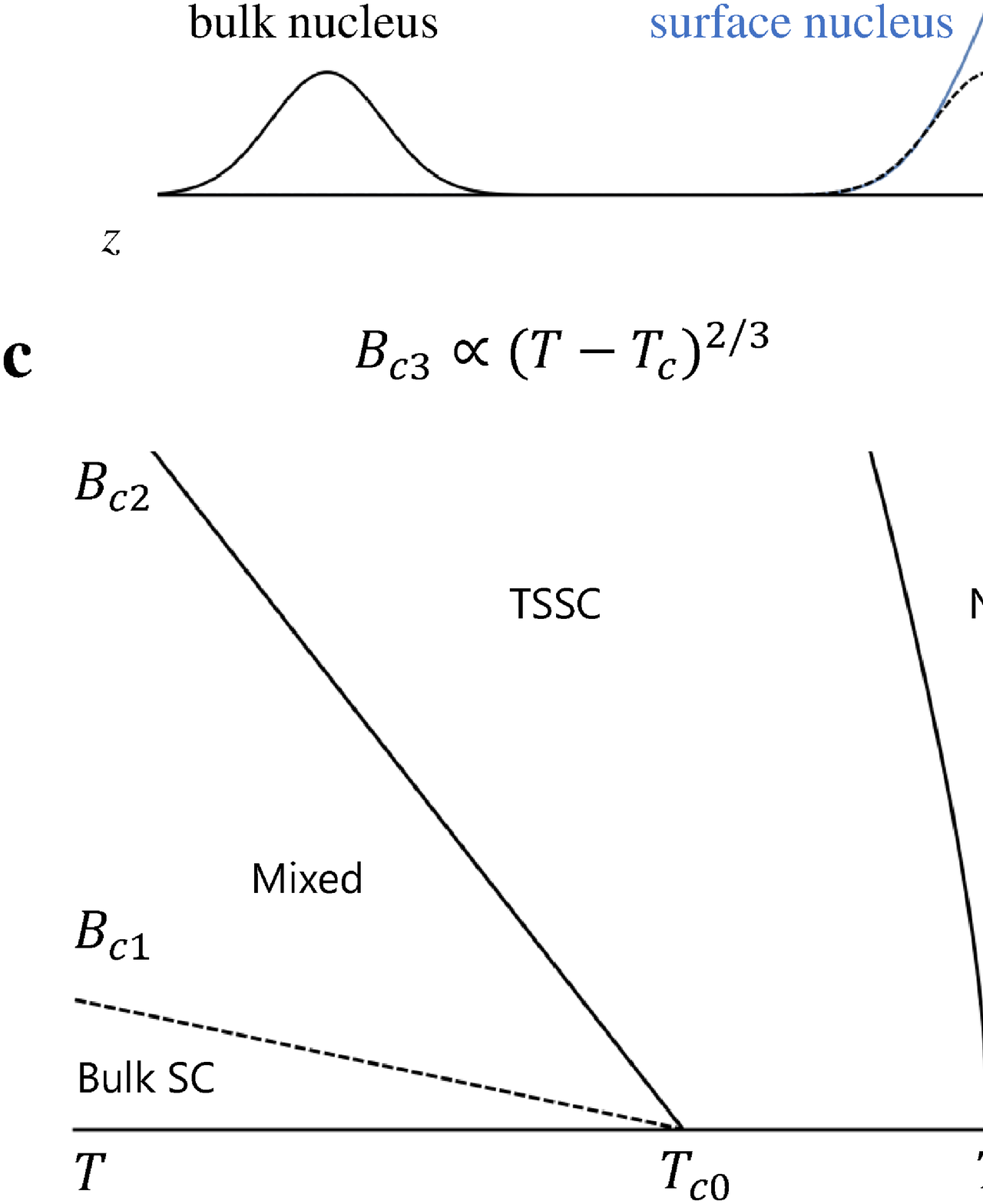}
\end{center}
\vspace{-0.5cm}
\noindent\textbf{Figure 2 $\mid$ Nucleation and phase diagrams under magnetic fields in the 3D topological superconductor}. \textbf{a, b}, Bulk and surface nucleation under magnetic fields. As the field increases, superconductivity first nucleates in the bulk (bulk nucleus) and then on the surface (surface nucleus). The spatial profiles of order parameters along the $z$-axis are plotted under the in-plane (\textbf{a}) and out-of-plane (\textbf{b}) magnetic fields. In \textbf{b}, bulk and surface nuclei are both localized in the $xy$-plane, but the bulk nucleus is uniform along the $z$-axis. \textbf{c, d}, Phase diagrams of temperature and magnetic field along in-plane (\textbf{c}) or out-of-plane (\textbf{d}) direction. The lower critical field $B_{c1}$, upper critical field $B_{c2}$, and nucleation critical field $B_{c3}$ separate the bulk uniform superconducting phase (Bulk SC), the mixed phase (Mixed), the topological surface superconducting phase (TSSC), and the non-superconducting normal phase (N). %In \textbf{c}, the unit of ticks in $B$-axis is $B_s=\Phi_0/(\pi l_s^2)$ (see maintext).

\newpage
\begin{center}
\includegraphics[width=\columnwidth]{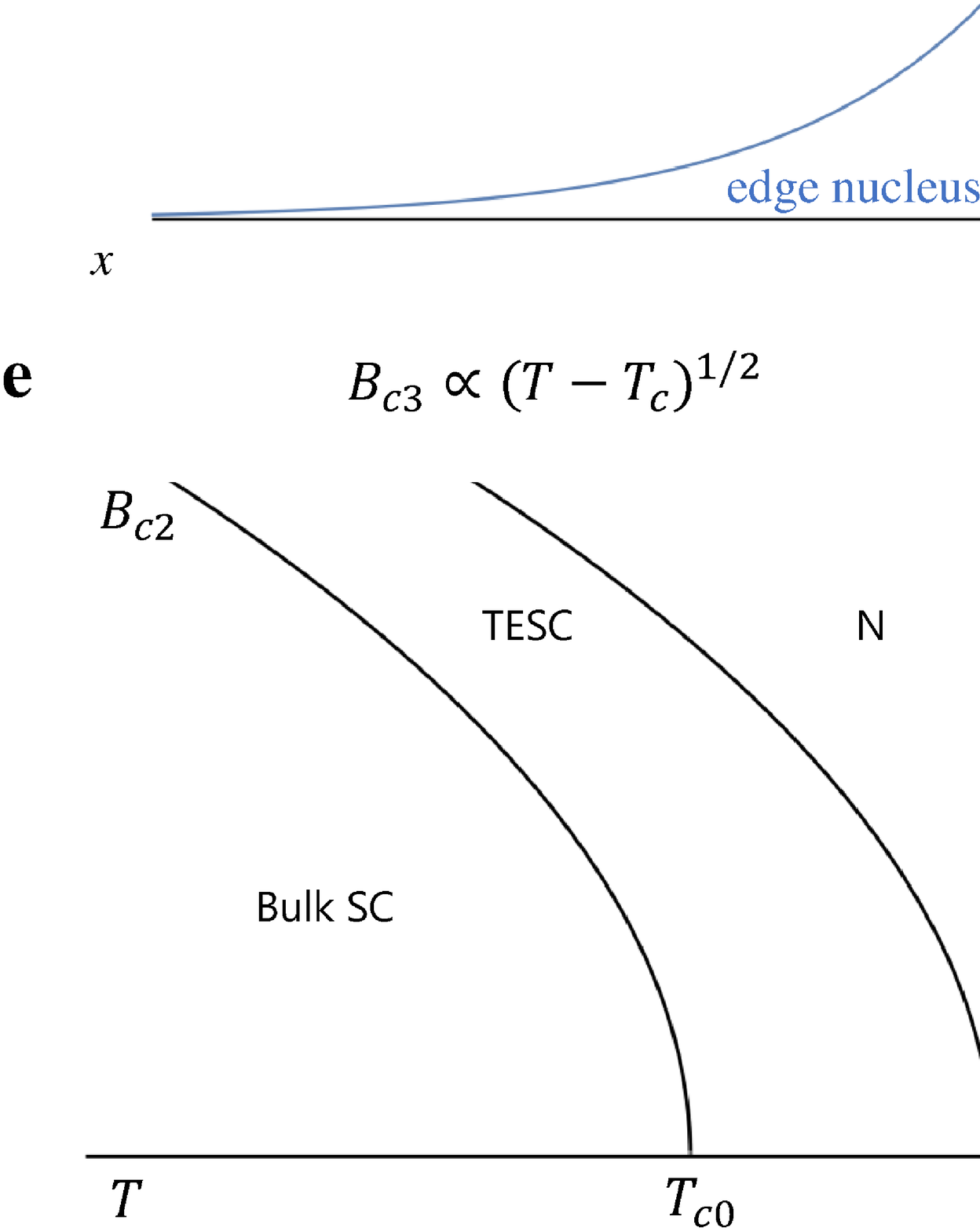}
\end{center}
\vspace{-0.5cm}
\noindent\textbf{Figure 3 $\mid$ Nucleation and phase diagrams in the 2D topological superconductor}. \textbf{a}, The 2D topological superconductor with bulk superconductivity (Bulk SC) in the semi-infinite region $x\leq 0$ and topological edge superconductivity (TESC) near the open edge $x=0$, and $\xi_0$ is the zero-temperature coherence length of the bulk. \textbf{b}, Edge nucleation at zero field. The blue and black lines correspond to the temperature regimes $T_{c0}<T<T_c$ and $T<T_{c0}$, respectively. \textbf{c-f}, Nucleation and phase diagrams under magnetic fields. Under the in-plane (\textbf{c, e}) and out-of-plane (\textbf{d, f}) fields, the order parameters along $x$-axis (\textbf{c, d}) and phase diagrams of temperature and field (\textbf{e, f}) are plotted. 
Under the in-plane fields (\textbf{c, e}), orbital effect is screened, hence the bulk nucleus is absent (\textbf{c}). As a result, the lower critical field $B_{c1}$, and the mixed phase (Mixed) are absent in the phase diagram (\textbf{e}).

\newpage
\begin{center}
\includegraphics[width=0.9\columnwidth]{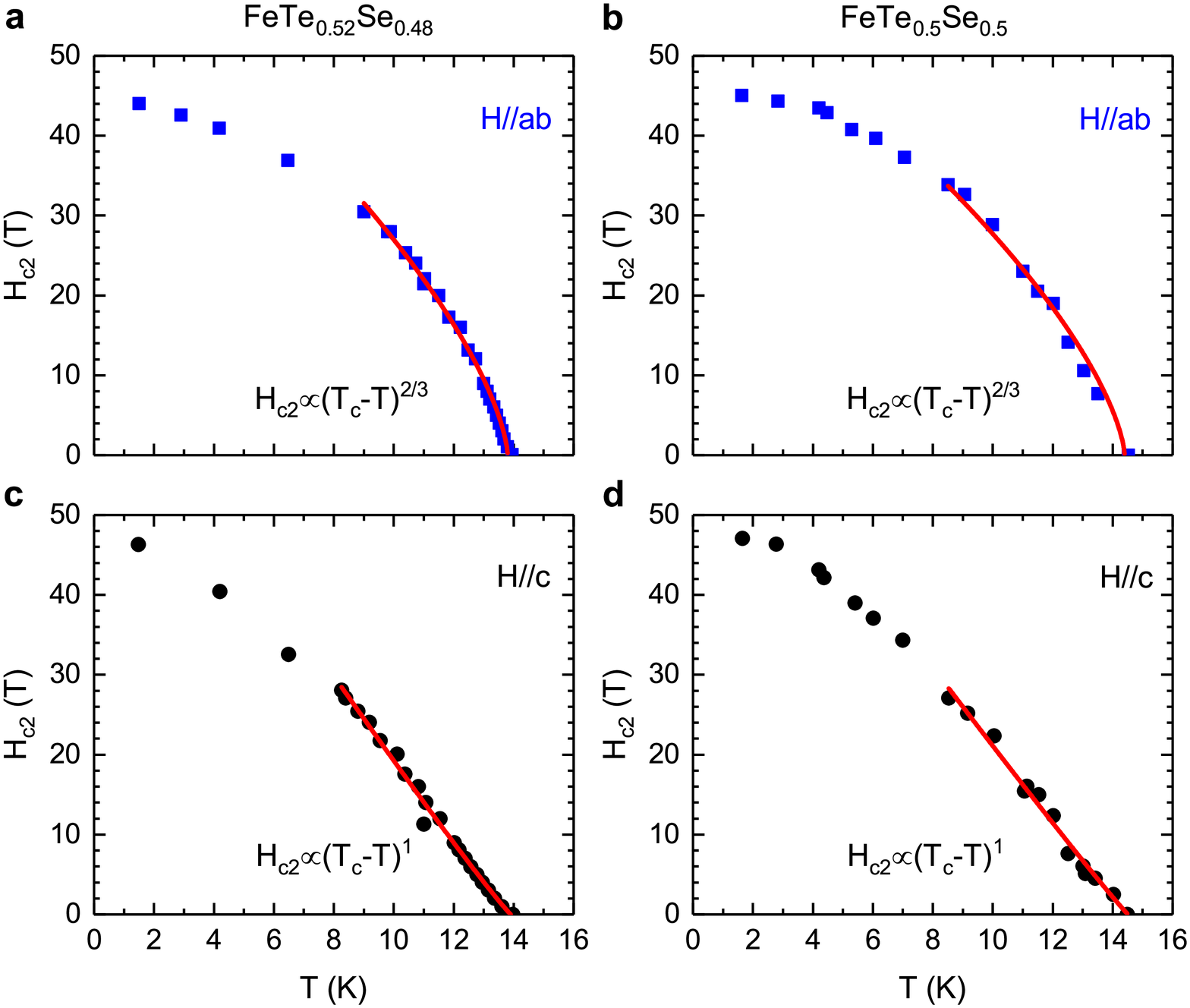}
\end{center}
\vspace{-0.5cm}
\noindent\textbf{Figure 4 $\mid$ Upper critical field behaviours of two iron-based superconductors FeTe$_{0.52}$Se$_{0.48}$ and FeTe$_{0.5}$Se$_{0.5}$ as a function of temperature for the magnetic field applied in the $ab$ plane and along the $c$ direction, respectively}. The symbols represent the experimental data points (squares for the $ab$ plane and circles for the $c$ direction) from the resistance and tunnel diode oscillator measurements on FeTe$_{0.52}$Se$_{0.48}$\cite{res} (\textbf{a} and \textbf{c}) and FeTe$_{0.5}$Se$_{0.5}$\cite{tdo} (\textbf{b} and \textbf{d}), respectively. The curves are the theoretical fitting results to the data points by using the critical exponent $\gamma$ of $2/3$ and $1$ for the $ab$ plane and $c$ axis for the topological superconductivity with the surface state given in Table 1.  

\newpage

\vspace{-0.5cm}
\noindent\textbf{Table 1 $\mid$ Critical exponents of the applied magnetic fields along the in-plane and out-of-plane directions in 3D and 2D topological and trivial superconductors}. The critical exponent $\gamma$ is defined by the nucleation critical field $B_{c3}\sim(T_c-T)^\gamma$ with $T_c$ being the onset critical temperature. In 3D, the plane is the open surface where topological surface states reside, while in 2D the plane is the superconductor itself. %Note that $\gamma=\frac{1}{2}\to\frac{2}{3}$ denotes two critical exponents $\frac{1}{2}$ and $\frac{2}{3}$ of the nucleation critical field $B_{c3}$ as the field increases from weak to intermediate values.

\begin{center}
	\begin{tabular}{c|cc|cc} \hline\hline
 Direction &3D Topological&3D Trivial& 2D Topological&2D Trivial \\
		\hline	
In-plane &${2}/{3}$&$1$&${1}/{2}$&${1}/{2}$\\
Out-of-plane &$1$&$1$&${2}/{3}$&$1$\\
		\hline	\hline
	\end{tabular}
\end{center}

\newpage

{\noindent\textbf{\large Methods}}

\noindent\textbf{Microscopic derivations.} To demonstrate the main steps of microscopic derivations, we consider the microscopic Hamiltonian at zero field with on-site attractive interaction among electrons
\be
    H=\int d^3\bm r d^3\bm s\left\{ c^{\dagger}(\bm r)\mathcal{H}(\bm r,\bm s)c(\bm s)\right\}
    -\frac{1}{2}g\int d^3\bm r \left\{c^{\dagger}(\bm r)c(\bm r)\right\}^2,
\ee
where $c=\{c_{\mu}\}^{\rm T}$, $c_{\mu}(\bm r)$ denotes an electron at site $\bm r$ with band index $\mu$, $\mathcal{H}(\bm r,\bm s)$ is the normal state Hamiltonian matrix of the topological material, and $g>0$ is the on-site attraction strength between electrons.
We would like to compute the $s$-wave pairing correlation $\Delta(\bm r)=g\langle c_{\uparrow}(\bm r)c_{\downarrow}(\bm r)\rangle$ in order to obtain the physics of superconductivity in this system, where $\langle O\rangle\equiv {\rm Tr}(O e^{-\frac{H}{k_{\rm B}T}})/Z$ denotes the thermodynamic average at temperature $T$, and $Z\equiv {\rm Tr}e^{-\frac{H}{k_{\rm B}T}}$ is the partition function.
To capture the long-range physics of $\Delta(\bm r)$, we expand the free energy $F\equiv {-{k_{\rm B}T}}\log Z$ within mean field
\be
    F=\int d^3\bm r\frac{|\Delta(\bm r)|^2}{g}-\int d^3\bm r d^3\bm s K(\bm r,\bm s)\Delta^*(\bm r)\Delta(\bm s),
\ee
where $K$ is the kernel in terms of sum over Mastubara frequency $\omega=(2n+1)\pi k_{\rm B}T$ $(n\in\mathbb{Z})$:
\be
    K(\bm r,\bm s)=k_{\rm B}T\sum_{\omega ab}
    \frac{\phi_{a}(\bm r)^{\dagger}\phi_{a}(\bm s)}{\xi_a-i\omega} 
    \left[\frac{\phi_{b}(\bm r)^{\dagger}\phi_{b}(\bm s)}{\xi_b-i\omega}\right]^{*},
\ee
and we introduce eigenstates $\mathcal{H}\phi_a=\xi_a\phi_a$, including both bulk states $\phi^{\rm B}$ and surface states $\phi^{\rm S}$.
Deep in the bulk, the kernel only involves bulk states and hence has full translation symmetry
\be
    K^{\rm B}(\bm r,\bm s)=k_{\rm B}T\sum_{\omega ab}
    \frac{\phi^{\rm B}_{a}(\bm r)^{\dagger}\phi^{\rm B}_{a}(\bm s)}{\xi_a-i\omega} 
    \left[\frac{\phi^{\rm B}_{b}(\bm r)^{\dagger}\phi^{\rm B}_{b}(\bm s)}{\xi_b-i\omega}\right]^{*}=K^{\rm B}(\bm r-\bm s).
\ee
The bulk critical temperature $T_{c0}$ and the bulk coherence length $\xi_0$ at zero temperature are determined by ($v_F$ is the bulk Fermi velocity)
\be
g\left.\int K^{\rm B}(\bm s)d^3\bm s\right|_{T=T_{c0}}=1,\quad \xi_0=0.18\frac{\hbar v_F}{k_{\rm B}T_{c0}}.
\ee

The introduction of the topological surface states on the open surface $z=0$ breaks the translation symmetry along the $z$ axis. 
On the scale $r\gg\xi_0$, the free energy can be rewritten as
\be
    F=\int_{z\leq 0} d^3\bm r\left\{\frac{|\nabla\psi|^2}{2m}+\mathcal{A}|\psi|^2\right\}+\sigma\int_{z=0} d^{2}\bm r |\psi|^2.
\ee
Here we introduce the superconducting order parameter $\psi$,
the temperature-dependent potential energy $\mathcal{A}$ and the temperature-independent surface energy $\sigma$ as follows\cite{deGennes}
\be
\psi(\bm r)={\sqrt{mL}}\Delta(\bm r),\quad
\mathcal{A}=\frac{1}{mL}\left\{\frac{1}{g}-\int K^{\rm B}(\bm r)d^3\bm r\right\},\quad
\sigma=\int_{z\leq 0}\frac{\Delta(\bm r)}{\Delta_0}\left\{1-\frac{N(\bm r)}{N_0}\right\}\frac{dz}{mgL},
\ee
where $N(\bm r)$ is the local density of states, $N_0=\underset{z \to -\infty}{\lim}N(\bm r)$ is the bulk density of states, and
\be
L=\frac{1}{3}\int K^{\rm B}(\bm r)|\bm r|^2 d^3\bm r,\quad 
\alpha=\frac{N_0}{mLT_{c0}},\quad
\Delta_0=\Delta(\bm r)|_{z=0}.
\ee

Without topological surface states, $N(\bm r)\equiv N_0$ and the surface energy is zero.
With topological surface states, $N(\bm r)>N_0$ and the surface energy is negative. Define the norm $||\phi||\equiv\sqrt{\phi^{\dagger}\phi}$, then $N(\bm r)=\sum_a||\phi_{a}(\bm r)||^2\delta(\xi_a)$ and $N_0=\sum_a||\phi_{a}^{\rm B}||^2\delta(\xi^{\rm B}_a)$. Hence
$N(\bm r)=N_0+N^{\rm S}(\bm r)>N_0$, where $N^{\rm S}(\bm r)\equiv\sum_a||\phi_{a}^{\rm S}(\bm r)||^2\delta(\xi^{\rm S}_a)$ is the local density of the topological surface states.
Hence
\be
\frac{1}{l_s}=\frac{2}{gL}\int_{z\leq 0}dz\frac{\Delta(\bm r)}{\Delta_0}\frac{N^{\rm S}(\bm r)}{N_0}.
\ee

Since $gL\sim\xi_0^2$, we expect $l_s\sim \xi_0^2/l_c$ where $l_c$ is the decay length of topological surface states\cite{TI}. When we take $l_c\sim$ 1 {\AA}, $\xi_0\sim$ 10 {\AA}, we have $l_s\sim$ 100 {\AA}. From the critical field data of FeTe$_{1-x}$Se$_{x}$ ($x\sim$0.5), we estimate $\xi_0$=21.68 {\AA} and $l_s$=276.65 {\AA}. The orders of magnitudes are consistent with our expectations.

\noindent\textbf{Ginzburg-Landau equation and de Gennes boundary condition.} 
By minimal coupling $-i\nabla\to\bm D=-i\nabla-2\pi\bm A/\Phi_0$, the derived Ginzburg-Landau free energy is then Eq. (\ref{eq_F}) of the maintext
\be
    F=\int_{z\leq 0} d^3\bm r\left\{\frac{1}{2m}\left|\bm D\psi\right|^2+\mathcal{A}|\psi|^2\right\}-\frac{1}{2ml_s}\int_{z=0} d^{2}\bm r |\psi|^2.
\ee
The variation of the Ginzburg-Landau free energy with respect to the order parameter reads
\be
\delta F=\int_{z\leq 0} d^3\bm r\left(\frac{\bm D^2}{2m}+\mathcal{A}\right)\psi\delta\psi^*
+ \frac{1}{2m}\int_{z=0} d^{2}\bm r(iD_z\psi-l_s^{-1}\psi)\delta\psi^*.
\ee

In the thermodynamic equilibrium, $\delta F=0$ for any $\delta\psi^{*}$. The variation in the bulk
leads to the linearized Ginzburg-Landau equation (GLE)
\bea
\left(\frac{1}{2m}\bm D^2+\mathcal{A}\right)\psi =0,
\eea
and the variation on the surface $z=0$ leads to the de Gennes boundary condition (dGBC)
\bea
\left.(iD_z\psi-l_{s}^{-1}\psi)\right|_{z=0}=0.
\eea
Finally, at infinity we find the von Neumann-type boundary condition
\be
\lim_{|\bm r|\to\infty}\bm D\psi=\bm 0.
\ee

\noindent\textbf{Nucleation at zero field in 3D topological superconductors.} Without external fields $\bm A=\bm 0$, near the phase transition the GLE is
\be
\left(-\frac{1}{2m}\nabla^2 +\mathcal{A}\right)\psi=0,
\ee
and the dGBC reads
\be
\left({\partial_z\psi}-{l_s}^{-1}\psi\right)_{z=0}=0.
\ee 
The order parameter is hence localized exponentially on the surface
\be
\psi =\exp(-{|z|}/{l_s}),
\ee
corresponding to the onset critical temperature
\be
T_c =T_{c0} +\frac{1}{2m\alpha l_s^2}.
\ee

\noindent\textbf{Nucleation under the in-plane fields in 3D topological superconductors.}
Without loss of generality, under a magnetic field $B\hat{\bm x}$ along the $x$ axis, we choose the Landau gauge $\bm A=(0,-Bz,0)$, then the GLE can be written as
\be
\left\{-\frac{1}{2m}\left[\partial_z^2+\partial_x^2-\left(i\partial_y-\frac{2\pi B}{\Phi_0}z\right)^2\right]+\mathcal{A}\right\}\psi=0,
\ee
and the dGBC is the same as the zero-field case since $A_z=0$:
\be
\left.(\partial_z\psi-l_{s}^{-1}\psi)\right|_{z=0}=0.
\ee
As plotted in Fig. 2\textbf{a} as the surface nucleus, the solution of the order parameter is
\be
\psi =D_{\nu}\left(-\sqrt{2}\frac{z+kl_B^2}{l_B}\right)e^{iky},\ \nu =\frac{1}{2}\left(\frac{B_{c2}}{B}-1\right),
\ee
where $l_B\equiv \sqrt{\Phi_0/2\pi B}$ is the magnetic length, and $D_{\nu}(\eta)$ is the parabolic cylinder function. 

Optimization of the critical field leads to 
$k=1/\xi$, and
the equation for the nucleation critical field $B_{c3}=B(T)$ is hence derived from dGBC
\bea\label{eq_1}
D'_{\nu}\left(-\frac{\sqrt{2}l_B}{\xi}\right)=\frac{-1}{\sqrt{2}}\frac{l_B}{l_s}D_{\nu}\left(-\frac{\sqrt{2}l_B}{\xi}\right).
\eea

At weak fields $B\ll B_s=\Phi_0/(\pi l_s^2)$, the solution to the above equation Eq. (\ref{eq_1}) is $B=\Phi_0/(\pi\xi l_s)\propto(T_c-T)^{1/2}$.
In the temperature regime $T_{c0}<T<T_c$, expanded to the leading order beyond exponential one can derive the order parameter
\be
\psi=\exp(-\frac{|z|}{l_s}-\frac{2}{3}\frac{|z|^3}{\xi^3l_s}+i\frac{y}{\xi}),
\ee
and hence the imhomogeneity of the order parameter is described by the length $\zeta=(\xi^2 l_s)^{1/3}$. It can be found that $B=0.525{\Phi_0}/{\zeta^2}\propto(T_c-T)^{2/3}$ in Eq. (\ref{eq_Bxy}) of the maintext is a good solution to Eq. (\ref{eq_1}) in the intermediate field regime $B\sim B_s$.

In the intermediate field regime, we can also introduce the post-Gaussian trial order parameter $\psi=\exp(-\frac{2}{3}z^3/\zeta^3-iry/l_B^2)$ with two variational parameters $r$ and $\zeta$, and minimize
\bea
\varepsilon(r,\zeta)\equiv{\int_0^{\infty}dz\left\{|\partial_z\psi|^2+\left|\left(\partial_y+i\frac{2\pi B}{\Phi_0}z\right)\psi\right|^2\right\}}\left/{\int_0^{\infty}dz|\psi|^2}\right. .
\eea
Then the nucleation critical field in the intermediate field regime can be worked out as
\be
B_{c3}=\frac{\Phi}{(\xi^2l_s)^{2/3}},\quad
\frac{\Phi}{\Phi_0}=\frac{1}{\sqrt[6]{6} \pi }\sqrt{\frac{\Gamma \left(\frac{1}{3}\right) \Gamma \left(\frac{2}{3}\right)}{\Gamma \left(\frac{1}{3}\right)-\Gamma \left(\frac{2}{3}\right)^2}}=0.489,
\ee
which is very close to the value $\Phi=0.525\Phi_0$ in Eq. (\ref{eq_Bxy}) of the maintext. 
And the other parameter is $r=\sqrt[3]{\frac{3}{4}}\frac{\Gamma \left({2}/{3}\right)}{\Gamma \left({1}/{3}\right)}\zeta$. 
The full phase diagram is shown in Fig. 2\textbf{c}, where the numerical solution $B=B_{c3}(T)$ to Eq. (\ref{eq_1}) is plotted.

\noindent\textbf{Nucleation under the out-of-plane fields in 3D topological superconductors.}
Under a magnetic field $B\hat{\bm z}$ along the $z$ axis, we choose the Landau gauge $\bm A=(0,Bx,0)$, then the GLE reads
\be
\left\{-\frac{1}{2m}\left[\partial_z^2+\partial_x^2-\left(i\partial_y+\frac{2\pi B}{\Phi_0}x\right)^2\right]+\mathcal{A}\right\}\psi=0,
\ee
and the dGBC is also the same as the zero-field case since $A_z=0$:
\be
\left.(\partial_z\psi-l_{s}^{-1}\psi)\right|_{z=0}=0.
\ee
So the solution of the order parameter is factorized as shown in Fig. 2\textbf{b} as the surface nucleus
\be
\psi =\exp(-\frac{|z|}{l_s})\exp\left(-\frac{|x-kl_B^2|^2}{2l_B^2}\right)e^{iky}.
\ee
We find the critical exponent for the out-of-plane fields is 1, the same as the conventional case\cite{tinkham,deGennes,Abrikosov}
\be
B_{c3,\perp}=\frac{\Phi_0}{2\pi\xi^2}\propto T_{c}-T.
\ee
The full phase diagram is shown in Fig. 2\textbf{d}.

\noindent\textbf{Nucleation under fields in 3D trivial superconductors.}
Without topological surface states, the superconductor is trivial, $N(\bm r)\equiv N_0$ and the surface energy is zero ($\sigma=0$). In this case, there is no nucleation at zero field and hence there is only one critical temperature $T_{c0}$. There will still be surface nucleation under finite magnetic fields as shown in Extended Data Fig. 1\textbf{a}, but the nucleation critical field is linear in temperature. The two critical fields have the following relation: $B_{c3}=1.69463B_{c2}$\cite{saint}. The full phase diagram is shown in Extended Data Fig. 1\textbf{b}.
In the regime $B_{c2}<B<B_{c3}$, the nucleated superconductivity becomes the \emph{trivial surface superconductivity}.

In summary, we find that in 3D the critical exponent of magnetic fields in the trivial case is 1 along any directions, while in the topological case, the critical exponent is 1 along the out-of-plane direction and 2/3 along the in-plane directions, where the plane is the open surface for topological states to reside. These results are listed in Table 1.

\noindent\textbf{Ginzburg-Landau free energy in 2D superconductors.} 
In a 2D superconductor, the surface energy reduces to the edge energy $\sigma'$, and
the Ginzburg-Landau free energy is
\be
    F=\int_{x\leq 0} dxdy\left\{\frac{1}{2m}\left|\bm D\psi\right|^2+\mathcal{A}|\psi|^2\right\}+\sigma'\int_{x=0} dy |\psi|^2,
\ee
where we assume the region of $x\leq 0$ is the 2D superconductor. When edge states exist along the open edge $x=0$, then $\sigma'=-({2ml_s})^{-1}<0$,
the GLE and dGBC in 2D are
\bea
\left(\frac{1}{2m}\bm D^2+\mathcal{A}\right)\psi =0,\quad
\left.(iD_x\psi-l_{s}^{-1}\psi)\right|_{x=0}=0.
\eea

\noindent\textbf{Nucleation at zero field in 2D topological superconductors.} Without external fields $\bm A=\bm 0$, near phase transition the GLE is
$
\left(-\nabla^2/2m+\mathcal{A}\right)\psi=0,
$
and the dGBC reads
$
\left.(\partial_x\psi-l_{s}^{-1}\psi)\right|_{x=0}=0.
$
The order parameter is hence localized exponentially on the edge
$
\psi =\exp(-{|x|}/{l_s}),
$
corresponding to the onset critical temperature
$
T_c =T_{c0} +({2m\alpha l_s^2})^{-1}.
$

\noindent\textbf{Nucleation under the in-plane fields in 2D topological superconductors.}
Under a magnetic field $B\hat{\bm n}$ in the $xy$ plane ($\hat{\bm n}\perp\hat{\bm z}$), we choose the Landau gauge $\bm A=Bz\hat{\bm n}\times\hat{\bm z}$, then $\bm A\equiv\bm 0$ since the superconductor is confined in the plane $z=0$. In other words, the orbital effect of magnetic fields can be neglected in 2D superconductors. 
But the potential term $\mathcal{A}$ is modified by the magnetic field via Zeeman effect, and a magnetic energy is introduced
\be\label{eq_F2}
\mathcal{A}\to \mathcal{A}'=\alpha(T-T_{c0})+\beta B^2.\quad (\beta>0)
\ee
Hence the GLE and the dGBC read
$\left(-\frac{1}{2m}\nabla^2 +\mathcal{A}+\beta B^2\right)\psi=0$, and $\left({\partial_x\psi}-{l_s}^{-1}\psi\right)_{x=0}=0$ respectively.
The solution is still localized exponentially on the edge $\psi =\exp(-{|x|}/{l_s})$ as shown in Fig. 3\textbf{a},
but we find the critical exponent for the in-plane fields is 1/2, the same as that in the conventional case of 2D superconductivity\cite{tinkham}
\be
B_{c3}=\sqrt{\alpha(T_c-T)/\beta}\propto (T_{c}-T)^{1/2}.
\ee
The full phase diagram of the in-plane magnetic field and temperature in the GL regime for 2D topological superconductors is shown in Fig. 3\textbf{c}. 
As field increases, the bulk superconductivity stays uniform until the upper critical field $B_{c2}=\sqrt{\alpha(T_{c0}-T)/\beta}$ where bulk superconductivity is killed.
In the regime $B_{c2}<B<B_{c3}$, nucleated superconductivity and topological states coexist along the edge, resulting in the \emph{topological edge superconductivity}.

\noindent\textbf{Nucleation under the out-of-plane fields in 2D topological superconductors.}
Under a magnetic field $B\hat{\bm z}$ along the $z$ axis, we choose the Landau gauge $\bm A=(0,Bx,0)$, then the GLE reads
\be
\left\{-\frac{1}{2m}\left[\partial_x^2-\left(i\partial_y+\frac{2\pi B}{\Phi_0}x\right)^2\right]+\mathcal{A}\right\}\psi=0,
\ee
and the dGBC reads
$
\left.(\partial_x\psi-l_{s}^{-1}\psi)\right|_{x=0}=0.
$
So the solution is as shown in Fig. 3\textbf{b}
\be
\psi =D_{\nu}\left(-\sqrt{2}\frac{x-kl_B^2}{l_B}\right)e^{iky},\ \nu =\frac{1}{2}\left(\frac{B_{c2}}{B}-1\right).
\ee
The full phase diagram of the out-of-plane magnetic field and temperature in the GL regime for 2D topological superconductors is shown in Fig. 3\textbf{d}. 
As field increases, the bulk superconductivity stays uniform until the lower critical field $B_{c1}$ where vortices start to nucleate in the plane. Then vortices proliferate and the bulk is in the mixed phase when $B_{c1}<B<B_{c2}$.
In the regime $B_{c2}<B<B_{c3}$, the nucleated superconductivity and topological states coexist along the edge, resulting in the \emph{topological edge superconductivity}.

\noindent\textbf{Nucleation under fields in 2D trivial superconductors.}
Without topological surface states, the superconductor is trivial, $N(\bm r)\equiv N_0$ and the edge energy is zero ($\sigma'=0$). In this case, there is no nucleation at zero field and hence there is only one critical temperature $T_{c0}$. 
There is no nucleation under in-plane magnetic fields as orbital effect is screened, and the Zeeman effect leads to $B_{c3}=B_{c2}=\sqrt{\alpha(T_{c0}-T)/\beta}$.
There is an edge nucleus under the out-of-plane magnetic fields similar to Extended Data Fig. 1\textbf{a}, and $B_{c3}=1.69463B_{c2}\propto T_{c0}-T$ still holds. The full phase diagram is shown in Extended Data Fig. 1\textbf{b}, except that in the regime $B_{c2}<B<B_{c3}$, the nucleated superconductivity is the \emph{trivial edge superconductivity}.

In summary, we find that in 2D, the critical exponent of magnetic fields in the trivial case is 1 along the out-of-plane direction and 1/2 along the in-plane directions, while in the topological case, the critical exponent is 2/3 along the out-of-plane direction and 1/2 along the in-plane directions, where the plane is the superconductor itself. These results are also listed in Table 1.

\noindent\textbf{Nucleation in superconducting heterostructures.}
When the region of $z>0$ is a metal or an insulator, a superconducting heterostructure is formed. 

Without topological states on the open surface or along the open edge, $N(\bm r)<N_0$ due to non-superconducting states near the open surface (edge) and the surface (edge) energy is positive ($\sigma>0$). {There is no surface (edge) nucleation} either at zero field or finite fields\cite{deGennes,Abrikosov}, hence the critical temperature, critical fields and the phase diagram of field and temperature are the same as those of the bulk.

With topological states on the open surface or along the open edge, the sign of the surface (edge) energy depends on the detailed competition between the topological states and the non-superconducting states near the open surface (edge). Our theory applies to the case of negative surface (edge) energy.

\newpage
\begin{center}
\includegraphics[width=0.9\columnwidth]{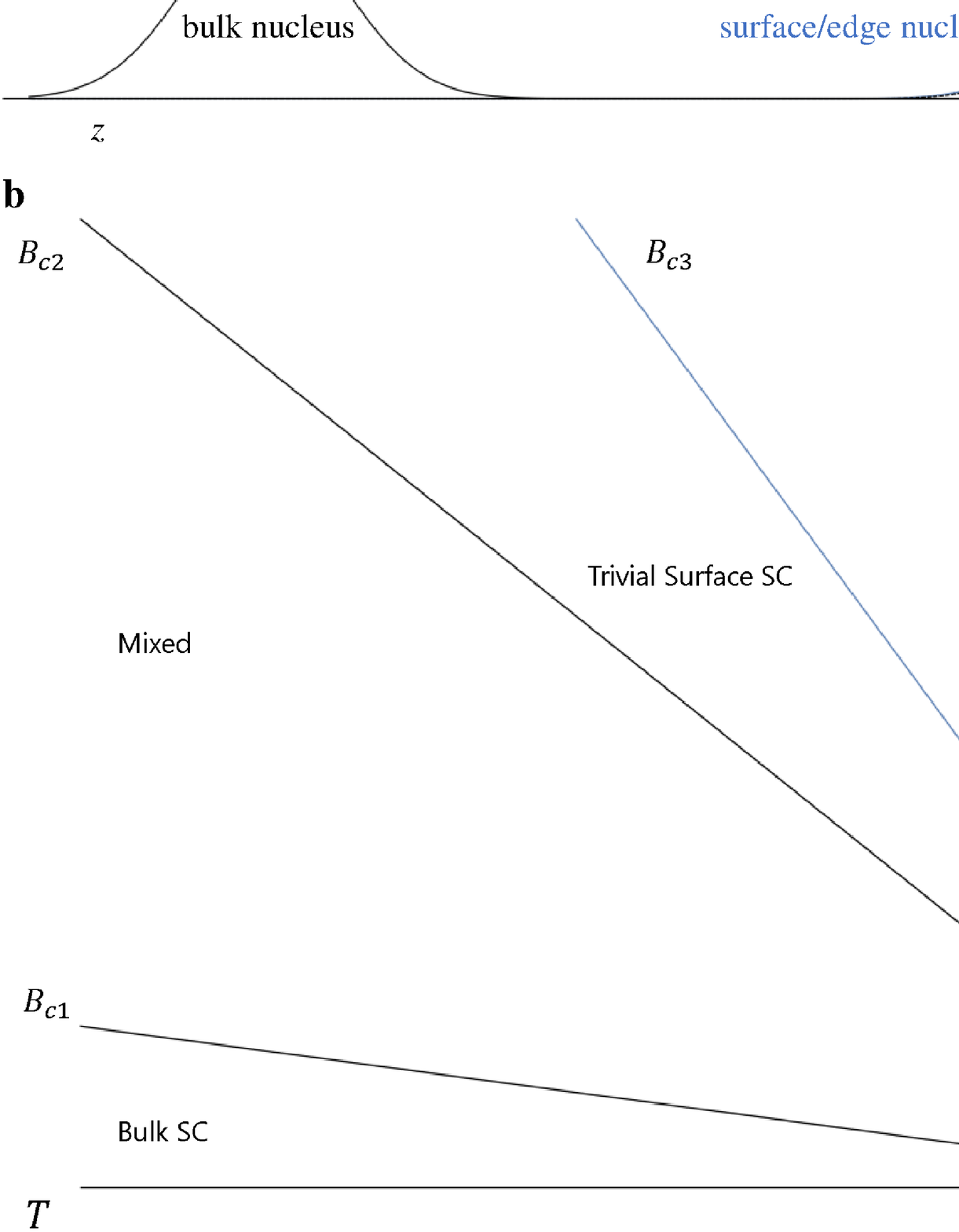}
\end{center}
\vspace{-0.5cm}
\noindent\textbf{Extended Data Fig. 1 $\mid$ Nucleation and phase diagram with zero surface energy (3D) or zero edge energy (2D)}. \textbf{a}, Bulk and surface nuclei with fields. \textbf{b}, Phase diagram of field and temperature. The lower critical field $B_{c1}$, the upper critical field $B_{c2}$, and the nucleation critical field $B_{c3}$ separate the bulk uniform superconducting phase (Bulk SC), the mixed phase (Mixed), the trivial surface superconducting phase (Trivial Surface SC), and the normal phase (Normal).

\end{document}